\documentclass[twocolumn,prl,showpacs,showkeys]{revtex4}

 \usepackage{graphicx}
 \usepackage{amsmath}

 \newcommand{\bE}{\mathbf{E}}
 \newcommand{\br}{\mathbf{r}}
 \newcommand{\bez}{\mathbf{e}_z}

\begin{document}
 \title{Supermodes of photonic crystal CCWs and multimode bistable switchings with uniform thresholds}
 \author{Weiqiang Ding}\email{wqding@hit.edu.cn}
 \author{Lixue Chen}
 \author{Shutian Liu}
 \affiliation{Applied Physics Department, Harbin Institute of Technology, Harbin, 150001, PRC}
 \date{\today}

 \begin{abstract}
 Photonic crystal (PC) coupled cavity waveguides (CCWs) with
 \textit{completely separated eigenfrequencies} (or supermodes) are investigated.
 Using a coupled mode theory, the properties of the
 supermodes, such as the electric field profiles, the eigenfrequencies
 (including the central frequencies and corresponding linewidths), and the quality factors
 (Qs) are expressed in very
 simple formulas, and these results agree well with those of ``exact'' numerical
 method. We also discuss the great differences between the supermodes and
 \textit{continuous modes}, which are described by tight binding (TB) theory. Then, the
 properties of supermodes are used to investigate their
 potential applications in multichannel (covering the whole C-band) bistable switchings,
 and bistable loops are obtained numerically. We
 also predict and verify that all the thresholds are uniform
 inherently.
  \\
 \end{abstract}

 \maketitle


 \section{Introduction}
 Photonic crystal (PC) \cite{pc:Eya87,pc:John87} coupled cavity waveguides (CCWs)
 \cite{ccow:ol,ccow:prl},
 which are formed by series of coupled point
 defects in otherwise perfect PCs, have been investigated intensively using tight
 binding (TB) theory in the last several
 years \cite{ccow:shaya:oe,ccow:shaya:preshg,ccow:shaya:pre02,ccow:jqe,ccow:JstQE}, and
 the propagation mechanism of hopping between
 neighboring cavities has been clearly
 understood \cite{ccow:hop}. Generally speaking, a periodic CCW possesses a
 continuous transmission band of
 $\Omega_0(1-\Delta\alpha/2+|\kappa|,1-\Delta\alpha/2+|\kappa|)$,
 where $\Omega_0$ is the eigenfrequency of an
 individual cavity and
 $\kappa$, $\Delta\alpha$ are two related overlap integrals
 \cite{ccow:ol,ccow:prl}. For convenience, the modes of the continuous band derived from
 TB theory are called \textit{TB-modes} in this paper.
 Many potential applications, such as
 broadband dropping \cite{ccow:banddrop}, broadband splitting \cite{ccow:spliter:apl,ccow:spliter:apl03},
 broadband optical switching and limiting \cite{ccow:ob:conti,ccow:OL:conti}, and
 ultrashort pulse transmission \cite{ccow:ultrashort} are all based on the
 continuous transmission bands (TB-modes).

 In practice, however, the number of the cavity $N$ is finite, and Born-von Karman periodic
 condition is used \cite{ccow:shaya:oe,ccow:JstQE} and $N$ \textit{discrete} modes are obtained,
 which satisfy the dispersion relation of
 \cite{ccow:shaya:oe,ccow:JstQE}:
 \begin{eqnarray}\label{eq:dispTB}
  & & \omega(k_m)= \Omega_0\left[1-\Delta\alpha/2+\kappa\cos(k_mR)\right]\\
  & & \mbox{with } k_m=2m\pi/L,\quad m=0,\cdots,N-1\nonumber
 \end{eqnarray}
 Where $R$ is the distance between two neighboring cavities, and $L$ is the total length
 of the system. When the coupling strength between the cavities are
 designed carefully, and some criteria are satisfied\cite{ccow:12D}, a quasi-flat spectrum with $N$ small dips
 formed by the $N$ discrete modes
 may be obtained\cite{ccow:12D}. Actually, the broadband operations mentioned above
 \cite{ccow:banddrop,ccow:spliter:apl,ccow:spliter:apl03,ccow:ob:conti,ccow:OL:conti}
 are all realized using the quasi-flat transmission band.

 \begin{figure}[ht]
 \centering
 \includegraphics[width=0.9\linewidth]{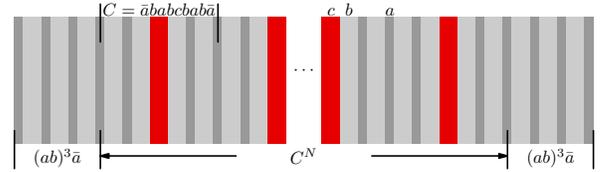}
 \caption{One dimensional photonic
 crystal CCW structure considered in this paper. The individual
 cavity of $C$ is represented by $C=\bar{a}babcbab\bar{a}$.
 Where $a$, $b$ and $c$ are three different kinds of medium layers with refractive indexes of
 $n_a$, $n_b$ and $n_c$, and physical thicknesses of $\lambda_0/(4n_a)$, $\lambda_0/(4n_b)$ and
 $\lambda_0/(2n_c)$, respectively. The layer $\bar{a}$ is the same as layer
 $a$ except that the thickness is $a/2$. $\lambda_0$ is a reference wavelength.}\label{fig:ccowab}
 \end{figure}

 On the other hand, one can also design a $N$-cavity CCW to make the $N$ modes \textit{separated
 completely}\cite{ccow:12D},
 (these separated modes are called supermodes in this paper and the reasons are given
 below), rather than to form a quasi-flat band.
 We find that the predictions of TB theory
 become inexact (as shown below) in this case. Although the supermodes of a $N$-cavity system
 have been noticed early\cite{ccow:12D}, the
 properties of them are not investigated intensively (such as the mode profiles and linewidths of
 each supermodes), which prevents them from wide application.

 According to the criterion developed in Ref. \cite{ccow:12D}, one dimensional (1D) CCW with
 discrete supermodes is designed, as shown in Fig. \ref{fig:ccowab}.
 Although some numerical methods, such as the finite difference time domain method (FDTD)
 \cite{fdtd:taflove}, transfer
 matrix \cite{tmm:linear,ccow:tmm} can be used to extract the properties we expected, we
 prefer to understand them from the viewpoint of physics and express them in analysis and
 simple formulas.

 In this paper, a general coupled mode theory is presented to analysis the supermodes of
 $N$-cavity CCW systems, as shown in Fig. \ref{fig:ccowab}. The eigenfrequencies, including
 both the central frequencies and the corresponding
 linewidths (and also the quality factors), as well as the mode profiles of the supermodes
 are obtained, and agree well with ``exact'' numerical results (obtained using standard transfer
 matrix method). Subsequently, these properties of supermodes are used directly to
 design a potential application of multichannel bistable switchings in the whole C-band.
 Using the results of coupled mode theory,
 we prove that the thresholds of the multichannel switchings are low and uniform when
 Kerr media are carefully introduced into the cavities.

 \section{Coupled mode theory analysis of supermodes}

 In the coupled mode theory presented in this paper, the electric field $\bE_{\omega}(\br)$ of
 the entire coupled cavity
 system is expressed as a linear superposition of the modes of the $N$ cavities:
 \begin{equation}\label{eq:sumEcmt}
 \bE_{\omega}(\br)=\sum_{n=1}^{N} A_n \bE_{\Omega}(\br-nR\bez)
 \end{equation}
 where $A_n$ $(n=1,\cdots,N)$ are complex coefficients that determine the relative phase
 and amplitude of the cavities. $\bE_{\omega}$ and $\bE_{\Omega}(\br-nR\bez)$ are the fields of
 the entire coupled system and those of individual cavities centered at $nR\bez$
 respectively. $\bez$ is the direction of the cavities being aligned.
 $\omega$ and $\Omega$ are the allowed frequency of the coupled system (is unknown now) and the
 frequency of an individual cavity (is already known).
 We normalize
 $\bE_{\Omega}(\br)$ to be unity according to $\int\epsilon^0(\br)\bE(\br)\cdot\bE(\br)=1$, with
 $\epsilon^0(\br)$ the dielectric function of an individual cavity.
 Eq. (\ref{eq:sumEcmt}) is very similar to the linear superposition field used in
 TB theory, which reads \cite{ccow:ol,ccow:prl}:
 \begin{equation}\label{eq:sumEtb}
 \bE_K(\br)=\sum_{n=1}^{\infty}B_n\bE_{\Omega}(\br-nR\bez)
 \end{equation}
 where the superposition coefficients are $B_n=E_0\exp(-inKR)$.
 However, one may find great differences between them.
 In TB theory (Eq.(\ref{eq:sumEtb})), the superposition coefficients are $B_n=E_0\exp(-inKR)$,
 which have the same modulars of $|E_0|$ for all the cases of $n$ from $1$ to $N$,
 and the relative phases between them are also
 determined. However, in Eq.(\ref{eq:sumEcmt}), the coefficients of $A_n$ are arbitrary complexes,
 and the amplitudes and phases of them may be greatly different for
 various of
 $n$. This is one of the most important differences between
 supermodes and TB-modes.

 Substituting Eq.(\ref{eq:sumEcmt}) into the simplified form of Maxwell's equations of:
 \begin{equation}
 \nabla\times\nabla\times\bE=\epsilon(\br)\frac{\omega_0^2}{c^2}\bE
 \end{equation}
 Then, we operate both sides of the resulting equation from left using the operator
 of $\int d\!\br \bE(\br-mR\bez)\cdot$, one can obtain a group of coupled equations:
 \begin{equation}\label{eq:Am}
 A_m+\sum_{m\neq n=1}^{N}C_{mn}A_n=0
 \end{equation}
 where the coefficients $C_{mn}$ are defined as

 \begin{gather}\label{eq:cmn}
 C_{mn} = \frac{\Omega^2\beta_{mn}-\omega^2\alpha_{mn}}{\Omega^2-(1+\Delta\alpha_m)\omega^2} \\
 \beta_{mn} = \int d\,\br\epsilon(\br-nR\bez)\bE(\br-mR\bez)\cdot\bE(\br-nR\bez) \\
 \alpha_{mn} = \int d\,\br\epsilon(\br)\bE(\br-mR\bez)\cdot\bE(\br-nR\bez) \\
 \Delta\alpha_m = \int d\,\br\Delta\epsilon \bE(\br-mR\bez)\cdot\bE(\br-mR\bez)\\
 (\mathrm{with\ }\Delta\epsilon_m(r) =\epsilon(\br)-\epsilon(\br-mR\br)\nonumber)
 \end{gather}

 \begin{figure}[ht]
 \centering
 \includegraphics[width=0.9\linewidth]{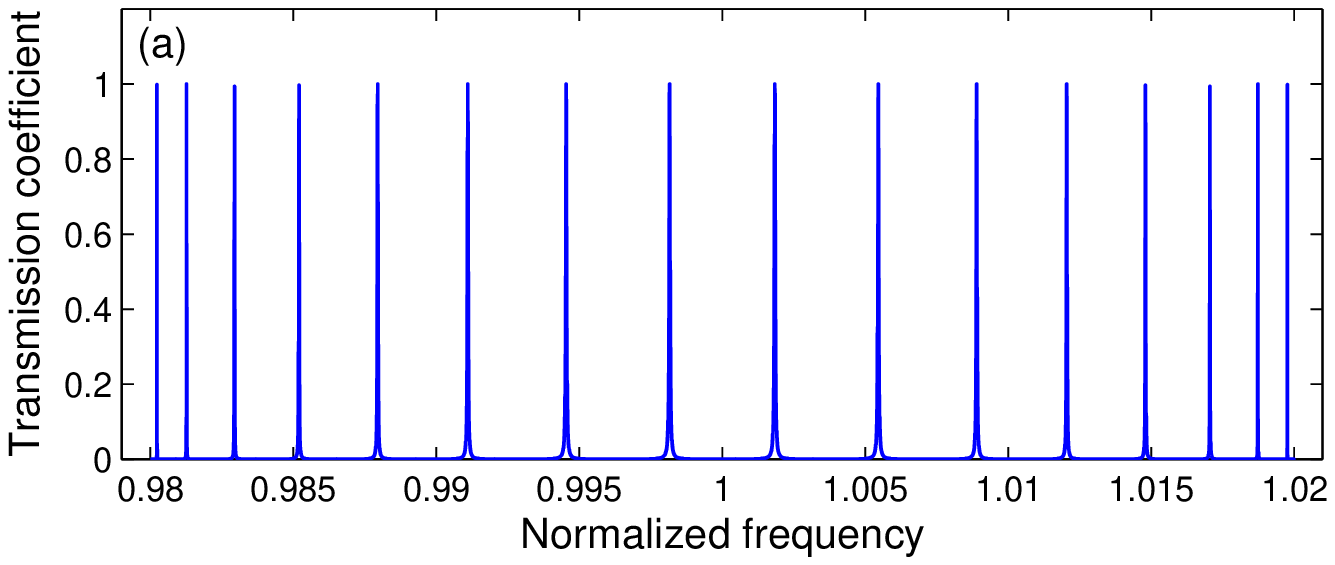}
 \includegraphics[width=0.9\linewidth]{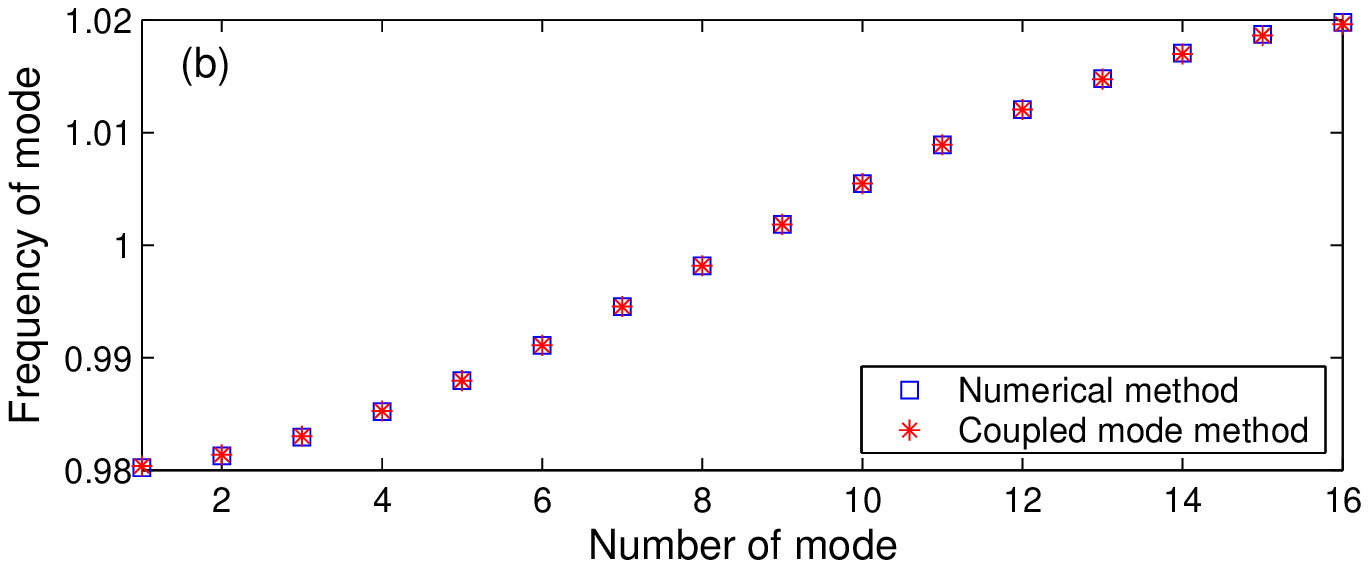}
 \includegraphics[width=0.9\linewidth]{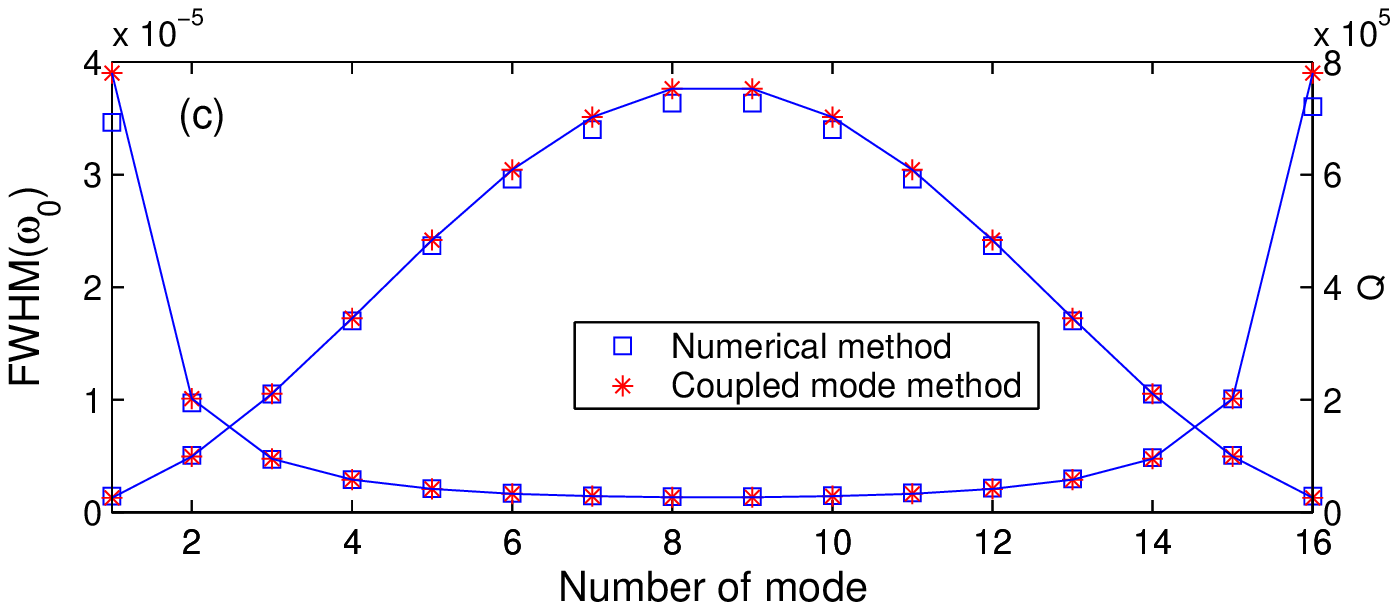}
 \caption{Comparison of the eigenfrequencies derived from the coupled mode
 theory and from numerical method for the structure of Fig. \ref{fig:ccowab}.
 The refractive indexes are $n_a=2$, $n_b=1$ and $n_c=2$
 respectively. Number of cavities is $N=16$.
 (a) Transmission spectrum calculated using one dimensional transfer matrix method.
 (b) Comparison of the central frequencies of numerical results (Fig. \ref{fig:TFQ}(a))
 with those derived from Eq. (\ref{eq:omegaL}) with parameters of
 $\alpha=0.00499$, $\beta=-0.01497$ and $\Delta\alpha=0$
 (See text for details).
 (c) Comparison the linewidths and quality factors of the supermodes derived from transfer matrix
 method (Fig. \ref{fig:TFQ}(a)) and coupled mode theory (Eq. (\ref{eq:Q})) with the
 parameter of $Q_0=2.6347\times10^4$.}
 \label{fig:TFQ}
 \end{figure}

 Eq. (\ref{eq:Am}) is very similar to the governing equations of coupled
 waveguide arrays \cite{book:yariv}, or
 phase-locked injection laser arrays \cite{cmt:apl,cmt:JQE,cmt:locklaser:ol}, where, the
 solutions are called supermodes of the waveguide (or laser) arrays. Similarly, we name the
 solutions of Eq.(\ref{eq:Am})
 as the supermodes of the CCW. Generally, Eq. (\ref{eq:Am}) can not be solved
 in closed form, however, it can be simplified and solved for some very special cases.
 For example, when only
 the nearest neighboring coupling are considered and the cavities are uniformly spaced.
 Then the coefficients are simplified to
 \begin{equation}
  C_{mn}= \frac{\Omega^2\beta-\omega^2\alpha}{\Omega^2-(1+\Delta\alpha)\omega^2}=
  \left\{%
\begin{array}{ll}
    C, & \hbox{$n=m\pm 1$;} \\
    0, & \hbox{others.} \\
\end{array}%
\right.
 \end{equation}
 Here, we have used the relations of $\beta_{m,m\pm1}=\beta$,
 $\alpha_{m,m\pm1}=\alpha$ and $\Delta\alpha_m=\Delta\alpha$.

 Then, using the same method as used in Ref.\cite{book:yariv,cmt:apl,cmt:JQE,cmt:locklaser:ol}
 and proper boundary conditions, one obtain
 the solutions of the Eq. (\ref{eq:Am}). Generally, there're $N$ solutions (supermodes)
 for a $N$-cavity CCW system. For the $L$th supermode, the linear superposition
 coefficients $A^L_n$ and $C^L$, respectively, are:
 \begin{eqnarray}
 A_n^L &=& A^L\sin(n\theta^L), \quad n=1,\cdots,N \label{eq:AnL}\\
 C^L &=& -\frac{1}{2\cos(\theta^L)} \label{eq:CL}\\
 \theta^L &=& \frac{L\pi}{N+1},\quad L=1,\cdots,N \label{eq:thetaL}
 \label{eq:CthetaL}
 \end{eqnarray}
 Where $A^L$ is a constant and is determined from the normalization condition of
 \begin{equation}\label{eq:normE}
 \int d\br\epsilon(\br)\left[\sum_{n=1}^{N} A_n \bE_{\Omega}(\br-n\bez)\right]\cdot
 \left[\sum_{n=1}^{N} A_n \bE_{\Omega}(\br-n\bez)\right]=1
 \end{equation}
 After a simple algebra process and using the normalization condition of the individual cavity
 modes, one can obtain the constant of $A^L$ is $A^L=\sqrt{2/(N+1)}$, which is
 independent of the mode number of $L$.

 According to Eq. (\ref{eq:cmn}) and Eq. (\ref{eq:CthetaL}), one can obtain
 the $N$ eigenfrequencies
 of the supermodes:
 \begin{equation}\label{eq:omegaL}
 \omega^L=\Omega\sqrt{\frac{C^L-\beta}{C^L-\alpha+C^L\Delta\alpha}}\quad L=1,2,\cdots,N
 \end{equation}

 Eq. (\ref{eq:Am}) and Eq. (\ref{eq:omegaL}) are the main results of the
 coupled mode theory discussed above. We want to point out that in
 Eq. (\ref{eq:CL}), $C^L$ tends to infinite when
 $\theta^L=m\pi+\pi/2$ ($m$ is an integer). For example, when $N=5$ and $L=3$.
 However, this does not mean
 that the coupled mode theory is invalid in this case. Because the parameter of $C^L$
 is a function of frequency of the supermode, $\alpha$ and $\beta$ (see Eq. (\ref{eq:cmn})),
 but not an physical quality. The values of physical qualities of $A^{L}_n$ (Eq.(\ref{eq:AnL}))
 and $\omega^L$ (Eq. (\ref{eq:omegaL}))
 are both finite and correct when compared with numerical results.

 Actually, the method shown above is not new, and it has been considered in Ref. \cite{ccow:prl}
 for the special cases of
 $N=2$ and $3$. When $N=2$, one can derive from Eq. (\ref{eq:CthetaL}) that $C^{1,2}=\mp1$,
 and then from
 Eq. (\ref{eq:omegaL}), one can obtain the frequencies of the supermodes and the superposition
 coefficients:
 \begin{equation}\label{eq:N2}
   \omega^{1,2}=\Omega\sqrt{\frac{1\pm\beta}{1+\Delta\alpha\pm\alpha}}, \quad
   A_{1,2}^1 = \frac{1}{\sqrt{2}}, \quad A_{1,2}^2=\pm \frac{1}{\sqrt{2}}
 \end{equation}
 Clearly, the superposition coefficients and the corresponding frequencies
 in Eq. (\ref{eq:N2}) are the same
 as the results of Eq. (2) in Ref. \cite{ccow:prl}. Similarly, one also can
 calculate the frequencies and fields for the case of $N=3$, which are also the
 same as the results in Ref. \cite{ccow:prl}.
 (In Eq. (3) of Ref. \cite{ccow:prl}, the $\Delta\alpha$ is regarded as negligible small).

 In order to show the power of the coupled mode theory discussed above, we consider
 a one dimensional CCW with
 a large number of cavities (i.e., $N=16$), as shown in Fig. \ref{fig:ccowab}.
 Using the standard transfer matrix method (TMM) \cite{tmm:linear}, one
 can easily obtain the transmission spectrum, as shown in Fig. \ref{fig:TFQ}(a). Using
 Eq. (\ref{eq:omegaL}), (\ref{eq:CL}) and (\ref{eq:thetaL}), one also can calculate
 the $16$ eigenfrequencies, and the results are shown in
 Fig. \ref{fig:TFQ}(b). Clearly, the theoretical results agree well with the numerical results.
 Here, the parameters are set to be
 $\alpha=0.00499$, $\beta=-0.01497$ and $\Delta\alpha=0$, which are obtained according to the
 following two steps: Firstly, we numerically obtain the two eigenfrequencies of
 $(\omega^{1,2}/\Omega_0)=0.990017,1.009983$
 for the case of $N=2$. Secondly using Eq. (\ref{eq:N2}), we calculate the above
 parameters of $\alpha$, $\beta$. Here we assumed that $\Delta\alpha$ is negligible compared
 to $\alpha$ and $\beta$. On the contrary, when the values of $\alpha$, $\beta$,
 $\Delta\alpha$ and $k_m$ are substituted into Eq. (\ref{eq:dispTB}), much larger
 errors of $\omega(k_m)$ from the numerical results are found. Therefore, the coupled
 mode theory gives more exact results than the TB theory in the case of discrete supermodes.

 The quality factors ($Q$'s) of the supermodes can also be derived from the
 results of coupled mode analysis. The $Q$'s of the $L$th supermode is defined as
 $Q^L=\omega^L W^L/\Delta W^L$, where $W^L=\frac{1}{2}\int_Vd\br\bE\cdot\mathbf{D}$ is
 the total energy stored
 in the coupled cavity system of the $L$th supermode, and $\Delta W^L=-dW^L/dt$ is the energy
 dissipated.
 According to the normalization condition of Eq. (\ref{eq:normE}),
 one can easily find that the total energy of all the $N$ supermodes are the same, i.e, $W^L=W_0$.
 For the CCWs composed of lossless media, the dissipation of energy of the $N$ supermodes are all due to the
 coupling out of the system through the $1st$ and $N$th cavities. Therefore, the energy
 dissipation of
 $\Delta W^L$ is proportional to the energy stored in the first and last cavities, i.e.,
 $\Delta W^L=\Delta W_0\sin^2(\theta^L)$. Here, we have used the results of Eq. (\ref{eq:AnL}).
 Then, one can find the quality factor $Q^L$ of the $L$th supermode:
 \begin{equation}\label{eq:Q}
 Q^L=\omega^L\frac{W^L}{\Delta W^L}=\omega^L\frac{W_0}{\Delta W_0\sin^2(\theta^L)}\approx\frac{Q_0}{\sin^2(\theta^L)}
 \end{equation}
 Here, $Q_0=\Omega_0W_0/\Delta W$, and we have used the the fact of
 $\omega^L\approx\Omega_0$. The results of Eq. (\ref{eq:Q}) have been shown in Fig.
 \ref{fig:TFQ}(c) with the parameter $Q_0=2.6347\times10^{4}$.

 On the other hand, we can also derive the $Q^L$ by finding the values of
 $Q^L=\omega^L/\Delta\omega_{fwhm}^L$ from the numerical results (Fig. \ref{fig:TFQ}(a)),
 where $\Delta\omega_{fwhm}^L$ is the full width at half maximum of the $L$th supermodes.
 The results are also shown in Fig. \ref{fig:TFQ}(c), and they agree well with the
 results of Eq. (\ref{eq:Q}).

\begin{figure}[th]
 \centering
 \includegraphics[width=0.9\linewidth]{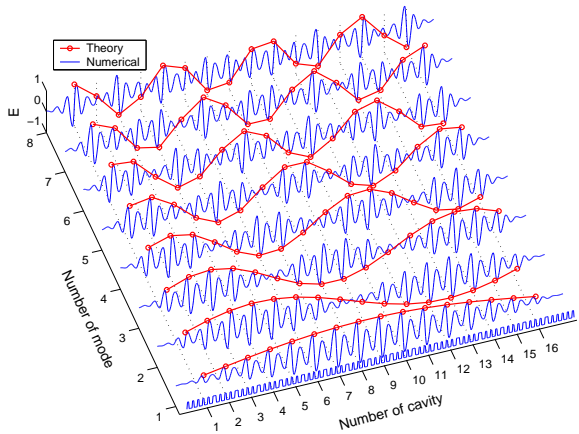}
 \caption{(Color online) The electric field profiles of supermodes calculated using
 transfer matrix method (blue lines) and the envelope of superposition coefficients (red lines with circles) shown in
 Eq. (\ref{eq:AnL}). The distribution of the
 refractive index is also shown for clarity.}\label{fig:Ein}
 \end{figure}

 Fig. \ref{fig:Ein} shows the field profiles of the first half supermodes of
 $\omega^{1,\cdots,8}$. For the case of clarity, the modes of $\omega^{9,\cdots,16}$,
 which are very similar to the modes of $\omega^{8,7,\cdots,1}$ respectively,
 are not shown. Clearly, the results of coupled mode theory of Eq. (\ref{eq:CthetaL}) are
 agree well with those of the transfer matrix method. The amplitudes of
 individual cavities of the $L$th supermodes lie on an envelope function of
 $\sin(\frac{xL\pi}{N+1})$. Therefore, the amplitude of some cavity modes may be tend
 to zero, while others reach to maxima. In the TB model, however, the amplitude of all
 the cavity modes are the same.

 From the discussion above, one can find that the coupled mode theory presented above
 describes the main characteristics of the supermodes very well, provided that
 the parameters of $\alpha$, $\beta$ and $Q_0$ are given.
 The supermode states of a CCW are very different with the states of TB-modes, which are
 described by the tight binding theory\cite{ccow:ol,ccow:prl}.
 For the TB-modes, the
 transmission band is determined by Eq. (\ref{eq:dispTB}), while in the supermodes, the central
 frequencies are determined by Eq. (\ref{eq:omegaL}).
 In the TB modes, the localizations in each cavities are the same, while in the supermodes,
 the localizations change greatly according to a simple sine
 function. In the next section, one of the possible applications of
 supermodes of multi-channel bistable switchings is proposed and
 analyzed using the results of coupled mode theory given above.

 \begin{figure}[ht]
 \centering
 \includegraphics[width=0.9\linewidth]{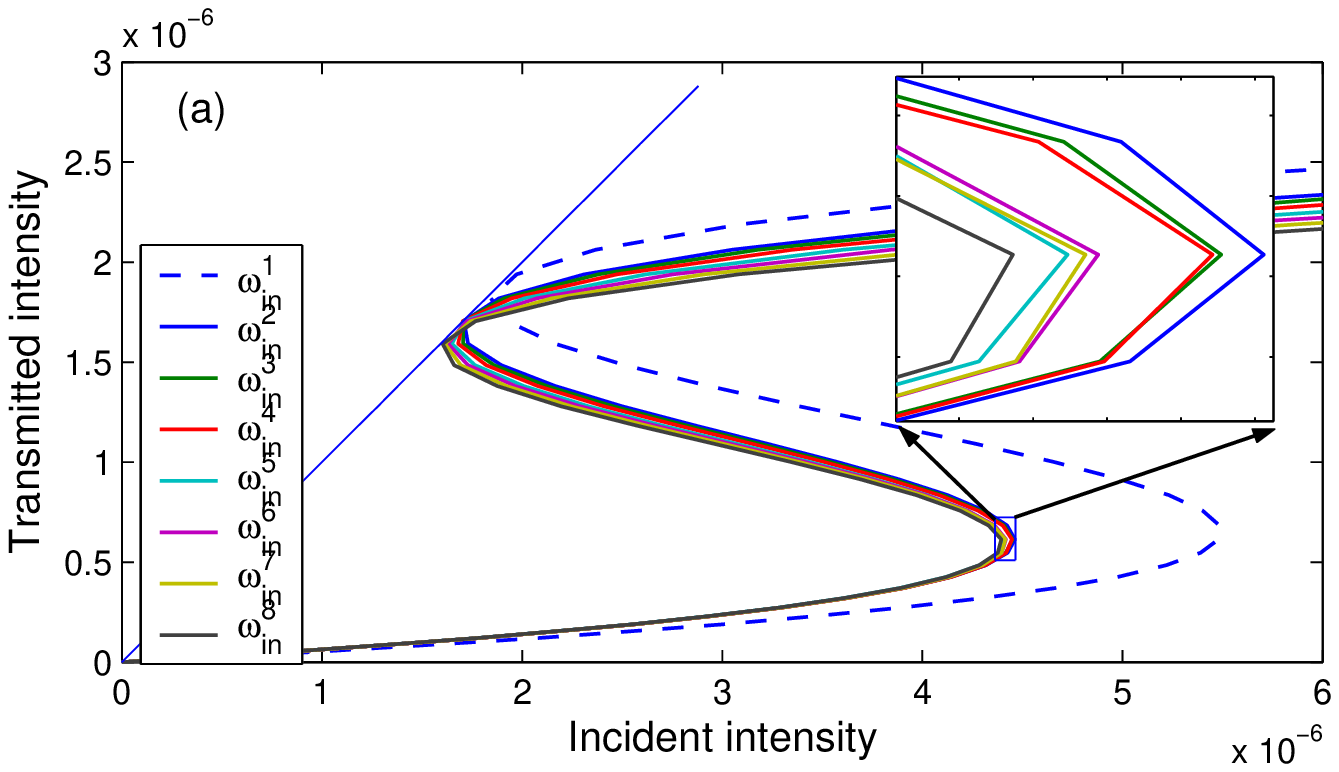}
 \includegraphics[width=0.9\linewidth]{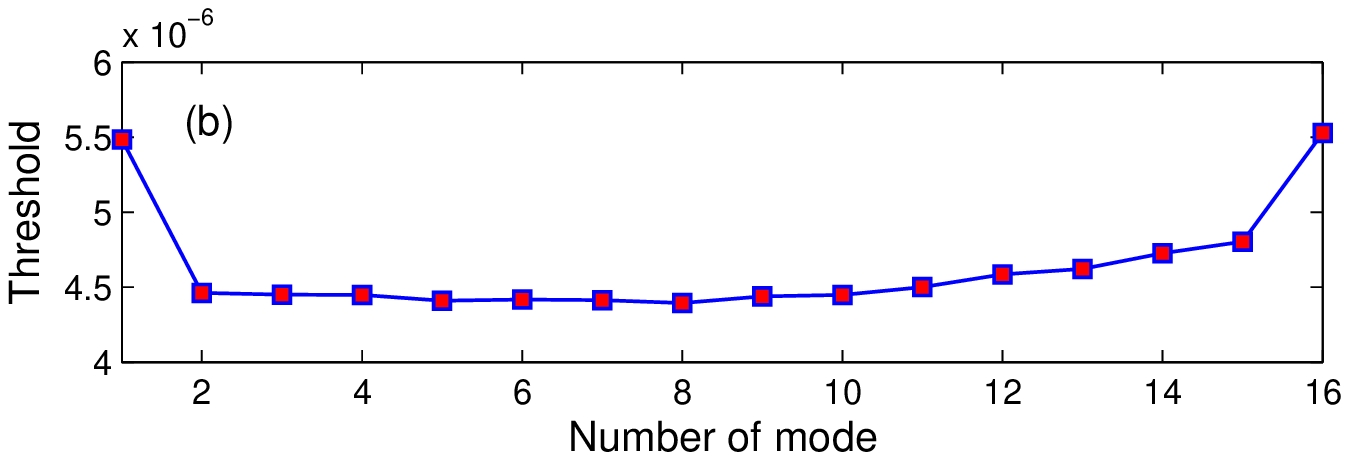}
 \caption{(Color online) (a) Bistable switching loops and the thresholds of
 the $16$-channel switchings.
 (a) Switching loops of $\omega_{in}^{1,\cdots,8}$, and those of $\omega_{in}^{9,\cdots,16}$ are
 not shown for clarity, which are similar to $\omega_{in}^{8,\cdots,1}$. Normalized unit of
 light intensity of $n_2I$ is used with $n_2$ the Kerr nonlinear coefficient and $I$ the light
 intensity. (b) the thresholds of the $16$-channel switchings.}
 \label{fig:ob}
 \end{figure}

 \section{Multi-channel bistable switchings with uniform thresholds}

 One of the potential applications of the supermodes is
 multichannel optical bistable (OB) switching, which is a key component in all
 optical information systems. The OB switching have been widely investigated
 in photonic crystals with Kerr defects
 \cite{ccow:ob:conti,ob:clx:oc02,ob:side,ob:dirct,ccow:ob:discr}, of which the refractive index
 changes with local light intensity. i.e., $n=n_0+n_2I$ with $n_0$ the linear
 refractive, $n_2$ the nonlinear Kerr coefficient and $I$ the local light intensity.
 However, most of the researches are focused on a
 single frequency operation \cite{ob:clx:oc02,ob:side,ob:dirct}.
 In a wavelength division multiplexer (WDM) system,
 there're always need a multi-channel switching for all the working channels. We find that
 the supermodes of CCWs are very suitable for this function \cite{ccow:ob:discr}.
 As an example, we investigate the systems of $16$-cavity CCW structures shown in Fig.
 \ref{fig:ccowab}.

 According to the OB switching theory, the shift of the eigenfrequencies with the
 changes of the dielectric constants of the cavities is a key factor for OB operation. Using
 the perturbation theory \cite{pertub:prb2002},
 one can find the shifts of eigenfrequencies of supermodes with the change of
 dielectric constant $\Delta\epsilon(\br)$:
 \begin{equation}\label{eq:purterb}
 \Delta\omega^L=-\frac{\omega^L}{2}\frac{\int\Delta\epsilon(\br)|\bE_{\omega^L}(\br)|^2}{\int\epsilon(\br)|\bE_{\omega^L}(\br)|^2}
 \end{equation}
 where $\bE=\sum A_n\bE(\br-n\bez)$ is the electric fields of supermodes of the unperturbed CCW.

 When only the $1$st and $N$th cavities are perturbed by the same amount of $\Delta\epsilon$,
 then using the relation of Eq. (\ref{eq:AnL}), (\ref{eq:normE}) and (\ref{eq:purterb}), one
 can derive that
 \begin{equation}
 \Delta\omega^L=-\frac{\omega^L}{2}\int_{C_1,C_N}\Delta\epsilon|\bE_{\omega^L}(\br)|^2d\br=\Delta\omega_0|\sin(\theta^L)|^2
 \end{equation}
 Where, $\Delta\omega_0=-[\omega^L\Delta\epsilon/(N+1)]\int_vd\!\br|\bE_{\Omega^L}(\br)+\bE_{\Omega^L}(\br-NR\bez)|^2$.
 On the other hand, according to Eq. (\ref{eq:Q}), we find that the FWHMs of the supermodes are
 $\omega^L_{fwhm}=(\omega^L/Q_0)\sin^2(\theta^L)$. Therefore,
 the shifts of the frequencies are proportional to their FWHMs respectively,
 i.e., $\Delta\omega^L=(\Delta\omega_0Q_0/\Omega_0)\omega^L_{fwhm}$.
 When the incident frequencies $\omega_{in}^L$ of the bistable switchings are tuned from the central
 frequencies according to $\omega_{in}^{L}=\omega^L-y\omega_{fwhm}^{L}$, the thresholds
 of the multi-channel bistable switching are expected to be uniform.

 Using the nonlinear transfer matrix method \cite{ntmm:apl92}, and setting $y=2$, we obtain
 the bistable
 switching loops, and the results are shown in Fig. \ref{fig:ob}.
 In Fig. \ref{fig:ob} we have used the normalized intensity of $n_2I$ \cite{ccow:ob:conti}.
 We can see that the thresholds of the switchings are almost the same. Except for the $1$st and
 $16$th channel (In practice, the $1$st and $N$th channels may be tuned slightly in order to
 switch them with approximately the same thresholds as other channels), the maximum of the
 thresholds is $4.8\times10^{-6}$ (the channel of $\omega_{in}^{15}$) and
 the minimum is $4.5\times10^{-6}$ (the channel of $\omega_{in}^{8}$),
 and the relative difference is about $6\%$.

 When the thicknesses of $a$, $b$ layers are $193.8nm$ and $387.5nm$ respectively, the
 central frequency of $\lambda_0$ is at about $1.55\mu m$, and the multi-channel frequencies
 cover the C-band of optical fiber communication entirely. For a Kerr nonlinear coefficient of
 $n_2=2\times10^{-13}cm^2/W$ (a value achievable in many nearly instantaneous
 nonlinear materials), the thresholds are about $4.6\times10^{-6}W/cm^2$, which is much
 smaller than those of the switchings studied before \cite{ccow:ob:conti,ob:clx:oc02}.

 \section{Conclusion}
 In summary, we have analyzed the supermodes of CCW systems, which are different from
 the TB-modes and are also important operation states of CCWs. Using the coupled
 mode theory, the eigenfrequencies,
 including the centers and the FWHMs,
 quality factors, and mode profiles of the supermodes are formulated in very simple forms.
 And they agree well with exact numerical results.
 We also discussed the great difference of the supermodes with the
 quasiflat TB-modes.
 We investigated one of the potential applications of the supermodes,
 which is a $16$-channel
 bistable switching with uniform thresholds covering the C-band of optical fiber
 communication. The results show that the thresholds are low and uniform.


 \end{document}